\documentclass{article}

\usepackage{blindtext}
\usepackage{hyperref}
\hypersetup{
    colorlinks=true,
    linkcolor=blue,
    filecolor=magenta,      
    urlcolor=cyan,
    pdftitle={Overleaf Example},
    pdfpagemode=FullScreen,
    }
\usepackage{amsfonts}
\usepackage{cite}
\usepackage{graphicx}
\usepackage{dcolumn}
\usepackage{bm}
\usepackage{amsmath}
\usepackage{amssymb}
\usepackage{hhline}
\usepackage[normalem]{ulem}
\usepackage{authblk}

\usepackage[numbers,sort,compress]{natbib}

\usepackage{color}

\date{}

\title{Polariton tweezers}
\author{N. I. Gribova$^{1,2}$\thanks{Email: gribova.ni@phystech.edu}, \,
E. S. Sedov$^{1,3,4}$, \,
A. V. Kavokin$^{1,2,3,5}$}
\affil{$^1$Abrikosov Center for Theoretical Physics, Moscow Institute of Physics and Technology, 9 Institutsky per, Dolgoprudny 141701, Russia}
\affil{$^2$Russian Quantum Center, 30 Bolshoi boulevard, Skolkovo, Moscow, 121205, Russia}
\affil{$^3$Spin Optics Laboratory, St. Petersburg State University, Ulyanovskaya 1, St. Petersburg 198504, Russia}
\affil{$^4$Stoletov Vladimir State University, 87 Gorky str., Vladimir 600000, Russia}
\affil{$^5$School of Science, Westlake University, 600 Dunyu Road, Hangzhou 310030, Zhejiang Province, China}

\begin{document}

\maketitle

\begin{abstract}
The non-invasive, deterministic manipulation of charge carriers remains a fundamental challenge in solid-state quantum technologies. Here, we demonstrate the concept of dynamic ``polariton tweezers'' capable of trapping and transporting electrons across a two-dimensional electron system via the long-range electrostatic potential of dipolar (d-) exciton-polaritons. By driving a semiconductor microcavity with a specifically engineered, non-collinear three-beam optical pump with TE-polarization, we synthesize a macroscopic polariton lattice with hexagonal symmetry. The spatially periodic field of this d-polariton condensate acts as a reconfigurable trapping landscape for the electrons. We show that introducing targeted frequency detunings between the continuous-wave pumps strictly breaks the spatiotemporal symmetry of the system, converting the stationary macroscopic interference pattern into a dynamically moving conveyor belt. By precisely customizing these frequency detunings, the entire lattice—along with the electrostatically trapped electrons—can be rigidly driven in any arbitrary in-plane direction. This precise phase-matching mechanism bypasses the need for a net kinetic momentum transfer, establishing a robust, optically controlled platform for the long-range routing of charged particles and topological defects in driven-dissipative quantum fluids.
\end{abstract}

%SSSSSSSSSSSSSSSSSS
%EEEEEEEEEEEEEEEEEE
%CCCCCCCCCCCCCCCCCC
\section*{Introduction}

Optical tweezers~\cite{ashkin1986observation,grier2003revolution} have become a versatile tool for the non-invasive manipulation of mesoscopic objects and isolated particles~\cite{moffitt2008recent}.
Their development has led to diverse specialized implementations, including holographic optical tweezers~\cite{dholakia2002optical, padgett2011holographic} for complex multi-trap arrays, plasmonic tweezers~\cite{juan2011plasmon, grigorenko2008nanometric} that confine light below the diffraction limit, near-field photonic crystal traps~\cite{rahmani2006optical}, and optomechanical manipulation platforms \cite{marago2013optical,giovanni2023roadmap}.
These techniques enable precise spatial control of dielectric particles, biological cells~\cite{zhong2013trapping}, and cold atoms~\cite{grimm2000optical}.
Extending optical trapping to the direct capture and continuous spatial transport of individual electrons, however, remains challenging because the gradient-force mechanism underlying conventional optical tweezers couples inefficiently to free charge carriers.
Their direct interaction with optical fields is instead governed by charge acceleration and ponderomotive forces, which generally require intense fields to produce substantial spatial manipulation \cite{RevModPhys79929,PhysRevX12031043}.

%FFFFFFFFFFFFFFFFF
%IIIIIIIIIIIIIIIII
%GGGGGGGGGGGGGGGGG
\begin{figure}
\centering
\includegraphics[width=0.6\linewidth]{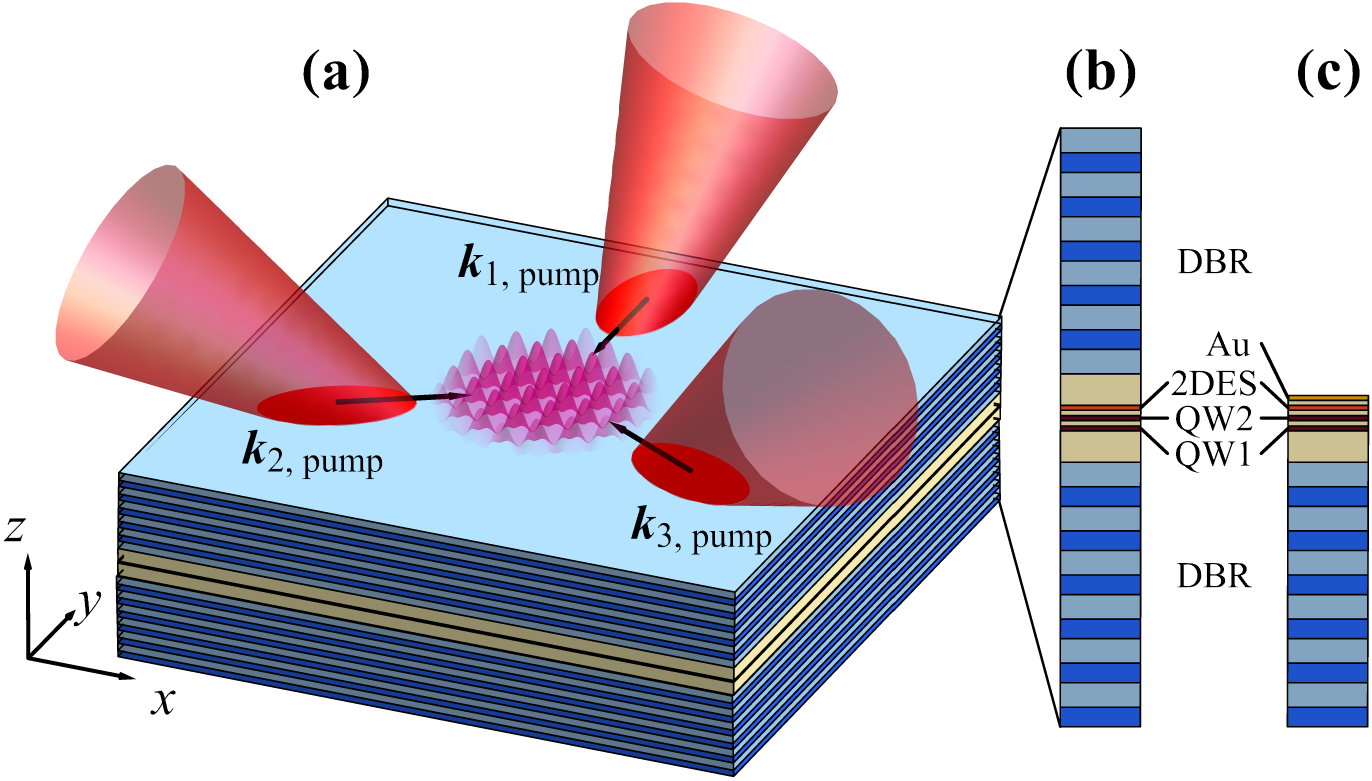}
\caption{\textbf{Schematic of the dipolariton microcavity and optical excitation geometry.}
\textbf{a}, Three-dimensional view of the system excited by three non-collinear coherent laser beams, depicted as semi-transparent cones with corresponding in-plane wavevectors $\mathbf{k}_{1\text{,pump}}$, $\mathbf{k}_{2\text{,pump}}$, and $\mathbf{k}_{3\text{,pump}}$.
Their overlap produces a spatially modulated dipolariton condensate, forming an interference lattice in the central region.
\textbf{b}, \textbf{c}, Enlarged cross-sectional views of two microcavity configurations, both incorporating tunnel-coupled quantum wells (QW1 and QW2) and an adjacent two-dimensional electron system (2DES). 
\textbf{b}, Conventional planar microcavity with a central cavity layer bounded by upper and lower DBRs.
\textbf{c}, Alternative Tamm-plasmon configuration in which the upper DBR is replaced by a thin gold layer
}
\label{fig:figure1}
\end{figure}

Parallel to the development of optical trapping~\cite{giovanni2023roadmap},  polaritonics has emerged as a robust solid-state platform for engineering light-matter interactions~\cite{kavokin2017microcavities,kasprzak2006bose,byrnes2014exciton,sanvitto2016road}.
Exciton-polaritons are hybrid light--matter quasiparticles formed in semiconductor microcavities through strong coupling between cavity photons and excitons hosted by active semiconductor layers, such as quantum wells (QWs), embedded within the cavity~\cite{kavokin2017microcavities}.
They can form macroscopically coherent states associated with Bose--Einstein condensation and exhibit strong optical nonlinearities~\cite{kasprzak2006bose,byrnes2014exciton,sanvitto2016road}.
The realization of dipolaritons (d-polaritons) in microcavities containing coupled QWs~\cite{christmann2011oriented,rosenberg2018strongly} has introduced an electrically tunable permanent dipole moment into polaritonic systems~\cite{nalitov2014voltage}.
As a result, d-polaritons combine the propagation characteristics inherited from their photonic component with long-range electrostatic interactions arising from spatially separated electron and hole components \cite{rosenberg2018strongly,kyriienko2014tunable}.

Polariton condensates can be spatially controlled using optically induced potentials~\cite{amo2010light,PhysRevB88041308,PhysRevB97235303,PhysRevB107045302}, surface acoustic waves~\cite{cerda2010polariton}, and engineered confinement geometries~\cite{SciRep134607,klembt2018exciton}.
Conversely, coupling between polaritons and charge carriers has been explored in several contexts, including condensate-assisted photocurrent generation and electron--polariton drag phenomena~\cite{myers2025drag,PhysRevB98045301,PhysRevX9041019}.
In the latter case, motion of the electronic subsystem transfers momentum to polaritons and modifies their transport.
However, to our knowledge, the use of a spatially structured polariton population itself as a moving electrostatic potential for the capture and transport of individual charge carriers has not been demonstrated.

A prerequisite for such a scheme is the ability to create and dynamically control spatially structured polariton density distributions.
This capability is already well established through the generation of engineered polariton lattices~\cite{jacqmin2014direct,klembt2018exciton,liew2008generation} and the guiding of polariton flows along engineered channels~\cite{wertz2010spontaneous,ACSPhot55095}.
Furthermore, frequency-detuned optical excitation has been used to create traveling optical lattices~\cite{del2024non} and dynamically controlled trapping potentials~\cite{SciAdv9eadd1299} for polariton condensates.
In a dipolariton system, such spatial modulation of the polariton density can be directly translated into an electrostatic potential landscape acting on nearby charge carriers.
Making this landscape dynamic therefore provides a route to optically controlled charge transport without relying on direct optical forces on the electron.

Here we theoretically propose a polariton tweezer architecture for the optical trapping and transport of individual electrons via the long-range electrostatic potential induced by a dynamically controlled d-polariton condensate.
Using three laser pump beams with non-collinear in-plane wavevectors, we generate a macroscopically occupied d-polariton state in a double-QW microcavity.
A controlled frequency detuning between the pump beams transforms the resulting stationary interference pattern into a traveling modulation of the d-polariton density.
Owing to the permanent dipole moment of the d-polaritons, this density modulation produces a moving electrostatic potential in an adjacent electronic layer.
We show that electrons initially confined within a potential minimum can remain entrained as the polariton lattice translates, providing a route to optically controlled charge transport and reconfigurable on-chip electronic routing.

%FFFFFFFFFFFFFFFFF
%IIIIIIIIIIIIIIIII
%GGGGGGGGGGGGGGGGG
\begin{figure*}[tb!] 
\centering
\includegraphics[width=0.99\linewidth]{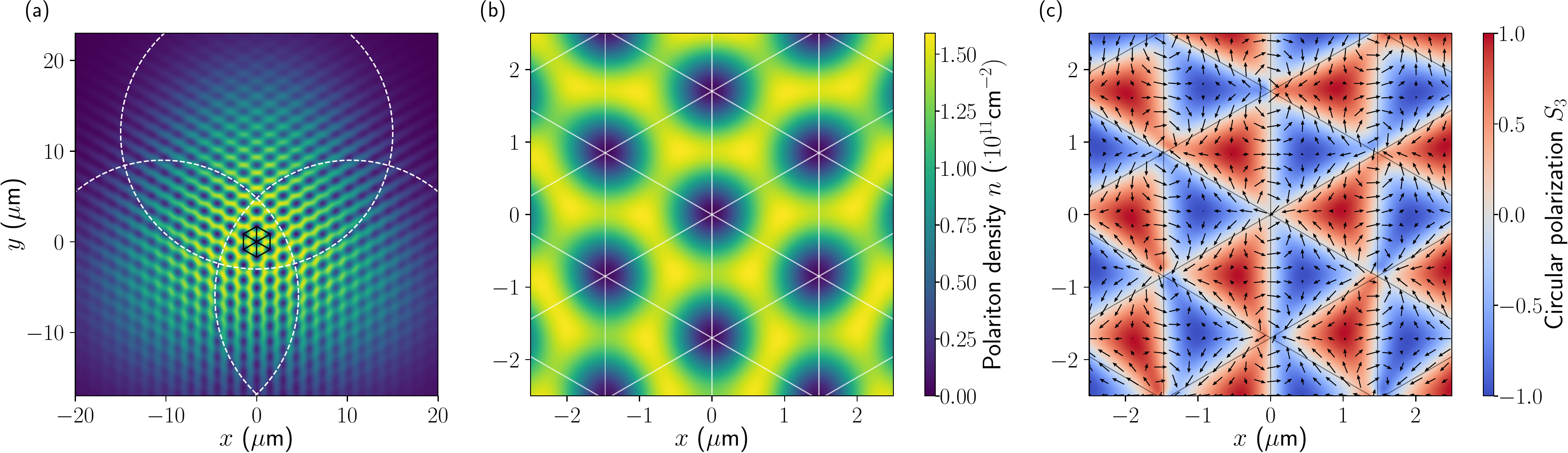}
\caption{ 
\textbf{Stationary polariton interference lattice.} 
\textbf{a}, Spatial distribution of the polariton density $n$.
White dashed circles indicate the positions and approximate spatial extent of the Gaussian pump spots. 
\textbf{b}, Enlarged view of the central interference lattice core.
Solid white lines outline its analytically determined hexagonal geometry. 
\textbf{c}, Spatial distribution of the circular polarization degree $S_3$ of the polariton condensate in the region shown in \textbf{b}.
The superimposed black arrows represents the in-plane Stokes components $(S_1, S_2)$, revealing local linear polarization textures and associated topological defects.
}
\label{fig:figure2}
\end{figure*}

%SSSSSSSSSSSSSSSSSS
%EEEEEEEEEEEEEEEEEE
%CCCCCCCCCCCCCCCCCC
\section*{Results}

%SSSSSSSSSSSSSSSSSS
%SSSSSSSSSSSSSSSSSS
%EEEEEEEEEEEEEEEEEE
%CCCCCCCCCCCCCCCCCC
\subsection*{Dipolariton microcavity platform}

The physical system under consideration is schematically shown in Fig.~\ref{fig:figure1}.
It consists of a planar semiconductor microcavity with a central cavity layer sandwiched between upper and lower distributed Bragg reflectors (DBRs), see  Fig.~\ref{fig:figure1}b.
The central cavity layer contains two tunnel-coupled quantum wells (QW1 and QW2), supporting spatially indirect excitons (IXs) with electrons and holes residing in different wells.
Such excitons are well established in structures with type-II band alignment~\cite{PhysRevB415283} and in coupled QWs with electric-field-controlled excitonic states~\cite{cristofolini2012coupling}.
Through electron tunnelling, IXs hybridize with optically active direct excitons, enabling strong coupling to the cavity photon mode and the formation of dipolaritons~\cite{cristofolini2012coupling}.
The resulting dipolaritons inherit a permanent out-of-plane electric dipole moment from their IX component, with its magnitude determined by the admixture of this component.

The system also includes a separate electronic layer hosting a two-dimensional electron system (2DES), as previously considered in electron--polariton hybrid structures~\cite{cotlect2016superconductivity}.
The electronic layer is positioned at distances $l_e$ and $l_h$ from the electron and hole layers of the IX, respectively.
An alternative Tamm-plasmon configuration (Fig.~\ref{fig:figure1}c), in which the upper DBR is replaced by a thin gold layer, may also provide a suitable platform, although metal-induced optical losses can reduce the cavity quality~\cite{wurdack2017observation,kaliteevski2009hybrid}.

The optical excitation geometry is schematically illustrated in Fig.~\ref{fig:figure1}a.
The microcavity is driven by three continuous wave Gaussian laser beams ($j=1,2,3$), with their spots arranged in a threefold symmetric configuration at a radial distance $R$ from the origin.
The beams have non-collinear in-plane wavevectors $\mathbf{k}_{j,\mathrm{pump}}$ with a characteristic magnitude $k_{\mathrm{pump}}$ and partially overlap in the central region.
Their coherent superposition produces a spatially periodic modulation of the dipolariton condensate density, forming an interference lattice.

The polariton condensate is described by the spinor wavefunction $\boldsymbol{\psi}(\mathbf{r},t)=(\psi_+,\psi_-)^\mathrm{T}$, whose components correspond to right- and left-circular polarizations.
Its dynamics is governed by the coupled driven-dissipative Gross--Pitaevskii equations, written in the frame rotating at the reference pump frequency $\omega_0$, chosen to be resonant with the TE-polarized mode:
\begin{equation}
\label{EqGPE}
i\hbar \frac{\partial \psi_\pm}{\partial t} = \biggl[ -\frac{\hbar^2}{2m^*} \nabla^2 - \hbar\omega_0 - i\frac{\hbar\gamma}{2} + \alpha_1 |\psi_\pm|^2  
+ \alpha_2 |\psi_\mp|^2 \biggr] \psi_\pm + \hat{\Omega}_{\pm} \psi_\mp + P_\pm(\mathbf{r}, t),
\end{equation}
where $m^*$ is the effective polariton mass in the cavity plane, $\gamma$ specifies the polariton decay rate, and $\alpha_1, \alpha_2$ denote the interaction constants for polaritons with the same and opposite circular polarizations, respectively.
The operators $\hat{\Omega}_{\pm} = \beta (\partial_x \mp i\partial_y)^2$ account for the TE--TM splitting of polariton modes~\cite{PhysRevB595082}, with strength $\beta$.

To characterize the spatial polarization distribution of the condensate, we introduce the normalized Stokes vector
$\mathbf{S}(\mathbf{r},t)=\boldsymbol{\psi}^\dagger\boldsymbol{\sigma}\boldsymbol{\psi}/n$,
where $\boldsymbol{\sigma}=(\sigma_x,\sigma_y,\sigma_z)$ denotes the vector of Pauli matrices and $n=\boldsymbol{\psi}^\dagger\boldsymbol{\psi}=|\psi_+|^2+|\psi_-|^2$ is the total polariton density.
The vector $\mathbf{S}$ specifies the local polarization state on the Poincar\'e sphere.

The coherent driving term in~\eqref{EqGPE} is represented as a superposition of three Gaussian pump beams with frequency detunings $\Delta\omega_j$ relative to the reference frequency $\omega_0$ and polarization-dependent phases $\theta_1=0$, $\theta_2=2\pi/3$, and $\theta_3=4\pi/3$:
\begin{align}
    P_\pm(\mathbf{r}, t) = 
    F_0 \sum_{j=1}^{3} \exp\left( -\frac{|\mathbf{r} - \mathbf{r}_j|^2}{\sigma^2} + i \mathbf{k}_{j,\mathrm{pump}} \cdot \mathbf{r} - i\Delta\omega_j t \mp i\theta_j \right),
\end{align}
where $F_0$ is the pump amplitude, $\sigma$ characterizes the spatial width of the Gaussian beams, and $\mathbf{r}_j$ and $\mathbf{k}_j$ denote their center positions and in-plane wavevectors, respectively.
The finite momentum spread of the Gaussian beams, of order $2/\sigma$, affects the spectral selectivity of resonant excitation and the relative contributions of the TE- and TM-polarized modes.
The opposite phase factors $\exp(\mp i\theta_j)$ are chosen to match the linear polarization of each pump beam to the TE-polarized mode, thereby favoring resonant excitation of the corresponding polariton branch.
Such a polarization configuration is consistent with the threefold symmetry of the excitation geometry, facilitating the formation and controlled manipulation of a symmetric interference lattice.

We consider three pump beams with independently adjustable frequency detunings $\Delta\omega_j$ ($j=1,2,3$) relative to the reference frequency $\omega_0=(1 + 2m^*\beta/\hbar^2)k_{\text{pump}}^2$.
Within the parabolic approximation, the TE--TM splitting modifies the dispersion of the selected polarization branch, and the wavevector magnitudes $K_j$ of the resonantly excited polariton modes are given by
\begin{equation}
K_j = \sqrt{ k_{\text{pump}}^2 +\frac{ 2m^*\Delta\omega_j/\hbar}{1 + 2m^*\beta/\hbar^2} }, \quad j \in \{1,2,3\}.
\end{equation}
The in-plane wavevectors are expressed as
$\mathbf{K}_1 = (0, -K_1)$, $\mathbf{K}_2 = (\sqrt{3}K_2/2, K_2/2)$, and $\mathbf{K}_3 = (-\sqrt{3}K_3/2, K_3/2)$.

%SSSSSSSSSSSSSSSSSS
%SSSSSSSSSSSSSSSSSS
%EEEEEEEEEEEEEEEEEE
%CCCCCCCCCCCCCCCCCC
\subsection*{Spatiotemporal control of polariton lattices}

The frequency detunings induce relative phase evolution between the three polariton fields, resulting in a traveling interference pattern.
The lattice drift velocity $\mathbf{V}=(V_x,V_y)^\mathrm{T}$ is determined by requiring the relative phases to remain time-independent in the frame moving with the lattice, $\mathbf{r}\mapsto\mathbf{r}+\mathbf{V}t$.
This condition gives $(\mathbf{K}_i-\mathbf{K}_j)\cdot\mathbf{V}=\Delta\omega_{ij}$, with $\Delta\omega_{ij} =\Delta\omega_i - \Delta\omega_j$.
For the wavevectors defined above, this yields
\begin{equation}
\mathbf{V} = \frac{1}{\mathcal{D}} \begin{pmatrix} 
\frac{1}{\sqrt{3}} \Big[ \Delta\omega_{13}(K_2 - K_3) + \Delta\omega_{23}(2K_1 + K_3) \Big] \\[10pt]
\Delta\omega_{23} K_3 - \Delta\omega_{13}(K_2 + K_3)
\end{pmatrix},
\end{equation}
where the denominator is $\mathcal{D} = K_1K_2 + K_1K_3 + K_2K_3$. 

When all frequency detunings vanish ($\Delta\omega_j=0$), the lattice drift velocity becomes zero ($\mathbf{V}=0$), resulting in a stationary interference pattern, as shown in Fig.~\ref{fig:figure2}(a,b).
In this case, all in-plane wavevectors have the same magnitude $K_0$. {The interference produces a regular hexagonal lattice where the horizontal spatial period is given by $a_0=4\pi/(\sqrt{3}K_0)$ (the distance between next-nearest neighbors along the $x$-axis), while the nearest-neighbor distance is $a_{\text{NN}}=4\pi/(3K_0)$}, as indicated by the white lines in Fig.~\ref{fig:figure2}(b).
The resulting density pattern exhibits $C_6$ symmetry and forms a hexagonal lattice.
It is accompanied by spatial textures in linear polarization and topological defects, as illustrated in Fig.~\ref{fig:figure2}(c).
In particular, the polarization texture contains alternating left- and right-half-vortices with fractional winding numbers $k,m=\pm 1/2$~\cite{rubo2007half} (see Supplementary Materials for details).

%FFFFFFFFFFFFFFFFF
%IIIIIIIIIIIIIIIII
%GGGGGGGGGGGGGGGGG
\begin{figure*}[htb!] 
\centering
\includegraphics[width=0.99\linewidth]{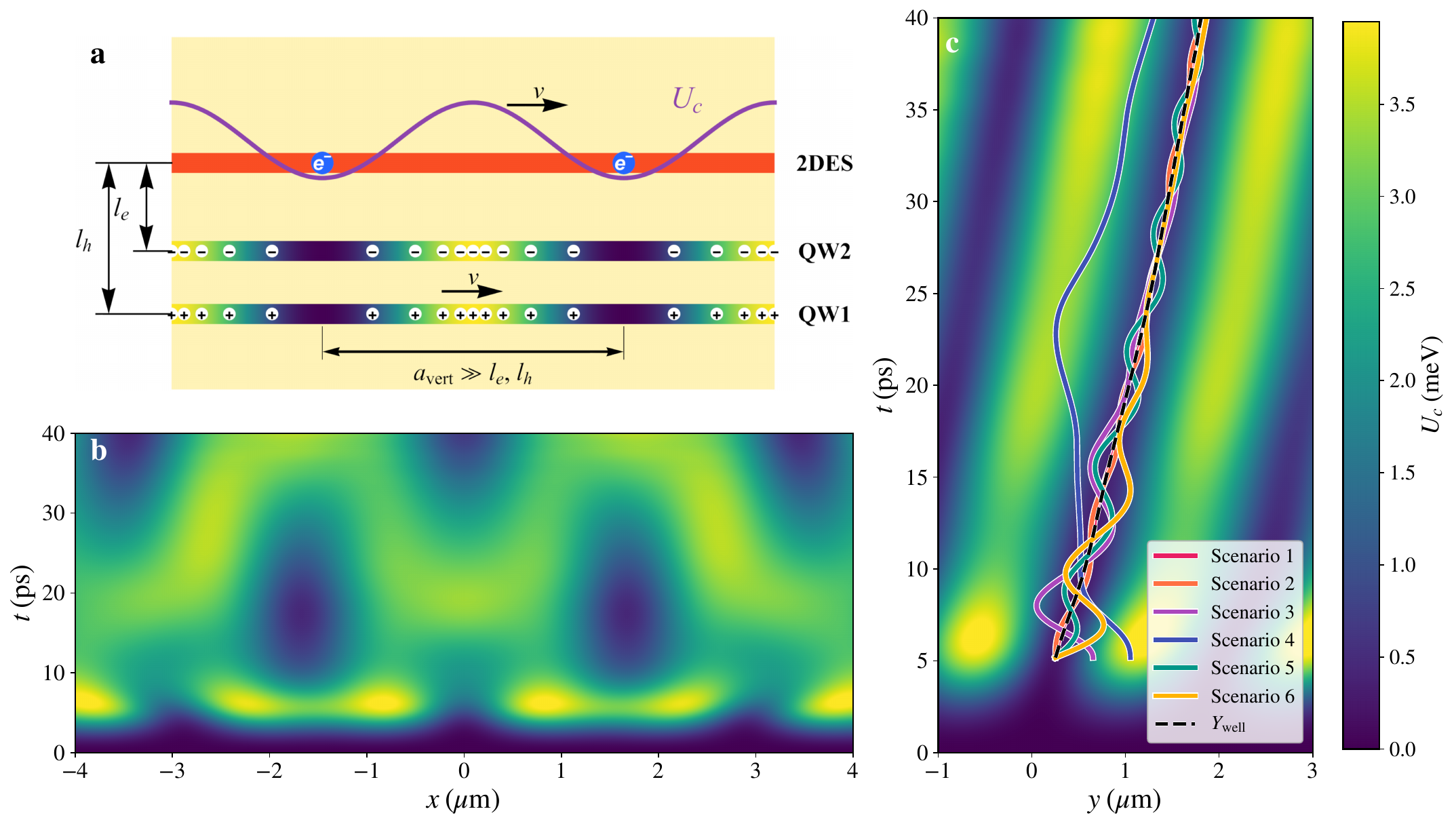}
\caption{ 
\textbf{Electron entrainment in a moving electrostatic lattice.}
\textbf{a}, Device schematic illustrating the electron trapping mechanism.
\textbf{b}, Spatiotemporal evolution of the polariton-induced electrostatic potential $U_c(x,t)$ along the cross-section at $y=0$.
\textbf{c}, Spatiotemporal evolution of $U_c(y,t)$ along the cross-section at $x=0$, overlaid with the mean longitudinal positions $\langle y\rangle(t)$ of the electron wave packets for six scenarios with different initial conditions.
The coloured trajectories illustrate how the electron wave packets follow the moving potential minima under different initial conditions.
}
\label{fig:figure3}
\end{figure*}

When one of the pump beams is frequency-detuned from the two others, $\Delta\omega_1\neq \Delta\omega_{2,3}=0$, the corresponding wavevectors satisfy $K_2=K_3=K_0= k_{\text{pump}}$, whereas $K_1$ varies with $\Delta\omega_1$.
The resulting interference lattice exhibits reduced spatial symmetry, from $C_6$ to $C_2$, and drifts along the $y$ axis with velocity $V_x=0$ and $V_y=-\Delta\omega_1/(K_1+K_0/2)$ (Fig.~\ref{fig:figure2}(b,c)).
The sign of $\Delta\omega_1$ determines the direction of motion, while the resulting variation of $K_1$ makes the drift speed a nonlinear function of the detuning~$\Delta\omega_1$.
{During this deformation, the horizontal spatial period $a_0=4\pi/(\sqrt{3}K_0)$ remains strictly unchanged. However, the nearest-neighbor distances become anisotropic: the vertical distance between adjacent density maxima along the $y$-axis scales as $a_{\text{vert}} = 2\pi/(K_1 + K_0/2)$, whereas the oblique nearest-neighbor distance varies according to $a_{\text{diag}} = \pi \sqrt{ 4/(3 K_0^2) + 1/(K_1 + K_0 / 2)^2 }$.}
The frequency detuning thus provides simultaneous control over the lattice drift velocity and its anisotropic deformation.
The TE--TM splitting strength $\beta$, determined by the microcavity structure, influences both quantities by modifying the relation between the frequency detuning and the polariton wavevector.

At a fixed reference pump frequency $\omega_0$, increasing the TE--TM splitting strength $\beta$ reduces the resonant wavevectors $K_0$ and $K_j$, thereby increasing both the drift speed $|V_y|$ and the characteristic lattice spacings $a_0$ and $a_{\mathrm{diag}}$.
Conversely, increasing the magnitude of the frequency detuning $\Delta\omega_1$ enhances the drift speed $|V_y|$, while positive (negative) detunings compress (expand) the diagonal lattice spacing $a_{\mathrm{diag}}$, leaving the transverse period $a_0$ unchanged.
These dependencies provide complementary means of controlling the motion and spatial geometry of the polariton interference lattice.

More generally, when multiple pump beams have nonzero frequency detunings, the interference pattern exhibits periodic temporal modulation if the independent frequency differences $\Delta\omega_{13}$ and $\Delta\omega_{23}$ are commensurate.
Otherwise, the temporal modulation is quasiperiodic.

%FFFFFFFFFFFFFFFFF
%IIIIIIIIIIIIIIIII
%GGGGGGGGGGGGGGGGG
\begin{figure*}[htb!] 
\centering
\includegraphics[width=0.99\linewidth]{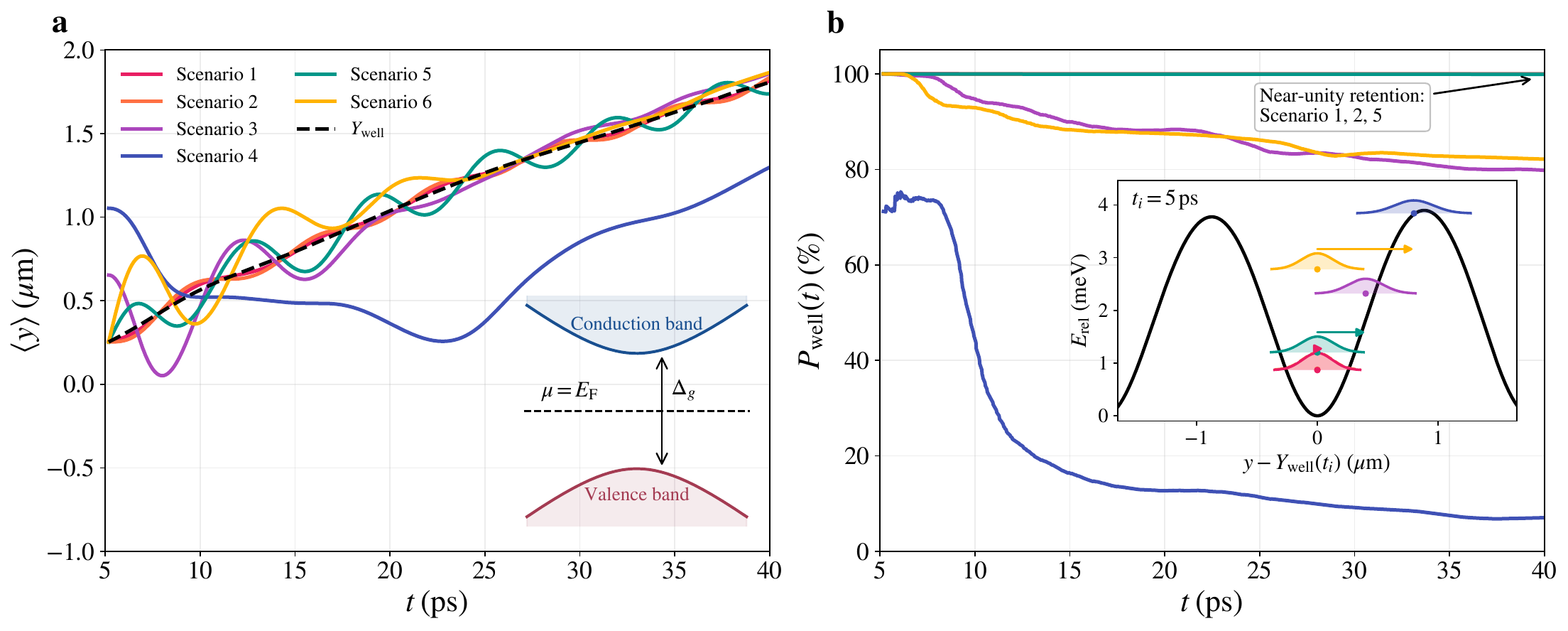}
\caption{ 
\textbf{Electron wave-packet transport under different initial conditions.}
\textbf{a}, Mean longitudinal position $\langle y\rangle(t)$ of the electron wave packets for six scenarios, compared with the trajectory $Y_{\mathrm{well}}$ of the moving potential minimum.
Scenarios 1, 2 and 5 closely follow the translating well, whereas Scenarios 3, 4 and 6 exhibit larger deviations.
The inset shows the gapped Dirac dispersion of graphene, with the chemical potential $\mu=E_{\mathrm F}$ located inside the bandgap $\Delta_g$.
\textbf{b}, Confinement probability $P_{\mathrm{well}}(t)$ within a circular region of radius $0.90~\mu$m centered on the moving potential minimum.
High probability retention is observed for Scenarios 1, 2, and 5.
The inset shows the confinement potential profile $U_c(y,t_i)-U_{\min}$ at the electron preparation time $t_i$, together with the initial electron probability distributions.
The profiles illustrate the initial positions and spatial widths of the wave packets, while the arrows indicate their initial propagation directions and relative speeds.
}
\label{fig:figure4}
\end{figure*}

%SSSSSSSSSSSSSSSSSS
%SSSSSSSSSSSSSSSSSS
%EEEEEEEEEEEEEEEEEE
%CCCCCCCCCCCCCCCCCC
\subsection*{Electron transport by dipolariton lattices}

The spatial separation of electrons and holes in QW1 and QW2 gives rise to permanent out-of-plane electric dipoles associated with the IX component of dipolaritons{, characterized by the Hopfield fraction $\eta_{\text{IX}}$}.
The spatial modulation of their density generates an electrostatic potential landscape $U_c(\mathbf{r},t)$ acting on electrons in the adjacent 2DES.
As illustrated in Fig.~\ref{fig:figure4}, the resulting Coulomb potential exhibits local minima that can serve as electron trapping sites.
The corresponding interaction kernel in momentum space is given by $\tilde{V}_0(\mathbf{q}) = \frac{e^2}{2\varepsilon_0\varepsilon \vert{}\mathbf{q}\vert{}} (e^{-\vert{}\mathbf{q}\vert{}l_e} - e^{-\vert{}\mathbf{q}\vert{}l_h})$, where $l_e$ and $l_h$ are the distances from the electron and hole layers of the IX to the 2DES, respectively, $e$ is the elementary charge, and $\varepsilon_0$ and $\varepsilon$ denote the vacuum and relative permittivities.
Since the characteristic spatial scale of the polariton lattice $a_0$ considerably exceeds the interlayer distances, {$a_0,a_{\text{vert}}, a_{\text{diag}} \gg l_e, l_h$}, the dominant Fourier components satisfy $\vert{}\mathbf{q}\vert{}l_{e,h} \ll 1$.
Expanding the exponentials to the first order, the momentum-independent interaction kernel is $\tilde{V}_0 \approx \frac{e^2 (l_h - l_e)}{2\varepsilon_0\varepsilon}$.
The resulting electrostatic potential is therefore approximately local and proportional to the {IX density $n_{\text{IX}}(\mathbf{r},t) = \eta_{\text{IX}} n(\mathbf{r},t)$}:
\begin{equation}
U_c(\mathbf{r},t) = \frac{e^2 (l_h - l_e)}{2\varepsilon_0\varepsilon} n_{\text{IX}}(\mathbf{r},t).
\end{equation}
For $l_h>l_e$, the potential is repulsive, and its minima coincide with the minima of the dipolariton density.
For the parameters specified in Methods, the calculated potential modulation reaches $3.2$~meV across the lattice (Fig.~\ref{fig:figure3}).

To examine electron confinement and transport, we solve the time-dependent Dirac equation for electron wave packets in the polariton-induced electrostatic potential (see Methods).
The 2DES is modeled as gapped graphene, and six initial wave-packet configurations are considered to assess the effects of initial position, velocity, and spatial width on electron transport relative to a moving potential minimum.
The same time-dependent electrostatic potential is used in all calculations, with the corresponding initial conditions specified in Table~\ref{tab:electron_regimes}.

%TTTTTTTTTTTTTTTTT
%AAAAAAAAAAAAAAAAA
%BBBBBBBBBBBBBBBBB
\begin{table}[t]
\centering
\caption{Initial conditions for the six electron transport scenarios.
The displacement $\Delta y_i$ is measured from the selected potential minimum along the direction of motion, $w_i$ is the width of the initial Gaussian envelope, and $u_y$ is the prescribed initial conduction-band group velocity.
Here $V_y$ denotes the lattice drift velocity.}
\label{tab:electron_regimes}
\begin{tabular}{lccc}
\hline
Scenario & $\Delta y_i$ ($\mu$m) & $w_i$ ($\mu$m) & $u_y$ ($\mu$m\,ps$^{-1}$)\\
\hline
1 (comoving) & 0 & 0.12 & $V_y$\\
2 (at rest) & 0 & 0.12 & 0\\
3 (displaced) & +0.40 & 0.14 & 0\\
4 (larger displacement) & +0.80 & 0.16 & 0\\
5 (finite velocity) & 0 & 0.13 & 0.25\\
6 (higher velocity) & 0 & 0.13 & 0.50\\
\hline
\end{tabular}
\end{table}

The trajectories of electron wave packets in Figs.~\ref{fig:figure3} and \ref{fig:figure4} show that, in Scenarion 1 and 2, the wave packets follow the translating potential minimum, despite their different initial velocities.
In Scenario 5, the packet also remains closely associated with the moving well, even when its initial speed exceeds the lattice drift speed.
An initial displacement of $0.40~\mu$m (Scenario 3) or higher initial velocity (Scenario 6) leads to oscillations around the moving minimum, whereas the larger displacement (Scenario 4) results in more pronounced deviations from the well trajectory and reduced confinement.

Confinement is quantified by the probability $P_{\mathrm{well}}(t)$ integrated over a circular region of radius $0.90~\mu$m centered on the moving potential minimum (Fig.~\ref{fig:figure4}b).
High probability retention is observed in Scenarios 1, 2 and 5, while Scenarios 3 and 6 retain a substantial confined component.
Scenario 4 exhibits more pronounced probability leakage from the well. 
These results demonstrate the transport of initially prepared electron wave packets without requiring exact matching of their initial velocities to the lattice drift velocity or electronic relaxation.

%SSSSSSSSSSSSSSSSSS
%EEEEEEEEEEEEEEEEEE
%CCCCCCCCCCCCCCCCCC
\section*{Discussion}  

We have demonstrated a concept for the dynamic spatial manipulation of  electrons, using the electrostatic landscape generated by a driven condensate of dipolaritons.
A defining advantage of this polaritonic tweezer architecture is its ultrafast transport capability, with estimated accessible lattice drift reaching  $0.2~\mu\text{m/ps}$.
In contrast to conventional solid-state charge-transport technologies, such as surface-acoustic-wave devices, limited by the speed of sound in the host material (e. g., $\sim 3\times10^{-3}~\mu\text{m/ps}$ in GaAs), or charge-coupled devices constrained by finite charge-transfer times and electrical clocking, including RC delays, the polaritonic lattice can propagate at substantially higher velocities.
The optically controlled motion of the polariton-induced potential thus offers a route to rapid in-plane transport of electrons in 2DES.

A key advantage of the proposed architecture lies in separating optical control from the physical mechanism of electron trapping.
Conventional optical tweezers rely on spatial gradients of optical intensity, whereas direct optical manipulation of free electrons is challenging and generally requires intense electromagnetic fields.
In contrast, the incident pump beams in our scheme generate and dynamically control a spatially structured dipolariton population, whose permanent electric dipoles produce a Coulomb potential acting on electrons in the adjacent 2DES.
Electron confinement is therefore governed by electrostatic rather than direct optical forces, while the position and motion of the trapping wells remain optically controllable.
Importantly, the spatial confinement of electrons is determined by the depth and curvature of the induced electrostatic potential rather than directly by the focusing of the incident optical fields.

Furthermore, the non-collinear pumping geometry allows the dipolariton lattice to operate as an optically controlled, two-dimensional  charge conveyor.
Unlike transport along predefined paths in gate-defined quantum dot arrays or surface-acoustic-wave channels, the lattice drift can be directed along an arbitrary in-plane orientation by adjusting the frequency detunings and resonant wavevectors of the incident beams.
This provides control over both the direction and speed of transport without modifying the physical geometry of the device.
The electrostatic potential depth can also be controlled through the dipolariton density and effective dipole moment, as well as the geometry of the hybrid heterostructure.
Meanwhile, the lattice spacing is governed by the pump wavevectors and polariton dispersion, including the effective mass and TE--TM splitting.

Looking forward, the ability to transport individual electrons along optically controlled trajectories without predefined physical channels offers opportunities for reconfigurable charge routing in solid-state devices.
Our results demonstrate that electron wave packets can follow the moving potential minima without requiring exact matching of their initial velocities to the lattice drift or electronic relaxation, although the degree of confinement depends on the initial conditions.
A complementary development in electron quantum optics is the on-demand generation of minimal-excitation single-electron states known as levitons~\cite{PhysRevLett97116403,Nature502659}, which provides a potential approach to preparing electronic excitations for subsequent controlled transport.
Extending our approach to spin-based logic and quantum information processing would require additional control over electron loading, release, and spin coherence during transport.
The proposed integration of dipolariton condensates with electronic systems thus provides a promising route toward optically programmable charge transport in integrated optoelectronic and spintronic architectures.

%SSSSSSSSSSSSSSSSSS
%EEEEEEEEEEEEEEEEEE
%CCCCCCCCCCCCCCCCCC
\section*{Methods}\label{methods}

%SSSSSSSSSSSSSSSSS
%SSSSSSSSSSSSSSSSS
%EEEEEEEEEEEEEEEEE
%CCCCCCCCCCCCCCCCC
\subsection*{Modeling of electron wave-packet dynamics}

The adjacent electronic layer is modeled as gapped graphene, described by a two-component wavefunction $\boldsymbol{\Phi}=(\Phi_A,\Phi_B)^\mathrm{T}$ corresponding to the two graphene sublattices.
The energy gap $\Delta_g$ accounts for sublattice-symmetry breaking, as observed in graphene/hBN heterostructures~\cite{giovannetti2007substrate,hunt2013massive,chen2014observation}.
Within the single-valley approximation, the electron dynamics in the polariton-induced potential $U_c(\mathbf r,t)$ is governed by the massive Dirac equation
\begin{equation}
i\hbar \partial _t \boldsymbol{\Phi}
=\left[\hat H_{D,0}+U_c(\mathbf r,t)\mathbb I_2\right]\boldsymbol{\Phi},
\end{equation}
where $\hat H_{D,0}= \hbar v_F(\sigma_x\hat{k}_x+\sigma_y\hat{k}_y) + \Delta_g \sigma_z/2$ is the free massive Dirac Hamiltonian, $\hat{\mathbf{k}}=-i\nabla$ is the wavevector operator, and $\sigma_{x,y,z}$ are the Pauli matrices.

For each initial condition, the electron wave packet is constructed from a Gaussian envelope $G_i(\mathbf r)=\exp[-|\mathbf r-\mathbf r_i|^2/4w_i^2+i\mathbf k_i\cdot\mathbf r]$, where $\mathbf r_i$, $w_i$, and $\mathbf k_i$ specify its initial position, spatial width, and central wavevector, respectively.
The initial state is projected onto the positive-energy branch of the free Dirac Hamiltonian:
\begin{equation}
\widetilde{\boldsymbol{\Phi}}_i(\mathbf k,t_i)
=\mathcal N_{\text{in}}\hat\Pi_+(\mathbf k)
(\widetilde G_i(\mathbf k),0)^\mathrm{T},
\end{equation}
where, $\widetilde G_i(\mathbf k)$ is the Fourier transform of $G_i(\mathbf r)$, and $\hat\Pi_+(\mathbf k)=[\mathbb I_2+H_{D,0}(\mathbf k)/\varepsilon_+(\mathbf k)]/2$, with $H_{D,0}(\mathbf k)$ denoting the momentum-space representation of $\hat H_{D,0}$ and $\varepsilon_+(\mathbf k)=\sqrt{(\Delta_g/2)^2+(\hbar v_F|\mathbf k|)^2}$ its positive eigenvalue.
The central wavevector is determined by the prescribed initial group velocity $\mathbf u_i$ through $\mathbf k_i=\Delta_g\mathbf u_i/[2\hbar v_F^2\sqrt{1-|\mathbf u_i|^2/v_F^2}]$, while $\mathcal N_{\text{in}}$ ensures normalization of the initial state.

The electron wave packet is initialized at $t_i=5$~ps, after the formation of the polariton lattice.
The six initial configurations differ in the displacement from a selected potential minimum, spatial width, and group velocity, allowing us to examine their influence on electron confinement and transport.
The resulting states are propagated under the time-dependent Dirac Hamiltonian. % using a Fourier split-operator method~\cite{mocken2008fft,bauke2011accelerating}.
The confinement probability is calculated as $P_{\mathrm{well}}(t)=\int_{|\mathbf r-\mathbf R_{\mathrm w}(t)|\leq a_e}\boldsymbol{\Phi}^{\dagger}\boldsymbol{\Phi}\,d^2\mathbf{r}$, where $a_e=0.90~\mu$m is the radius of the integration region and $\mathbf R_{\mathrm w}(t)$ denotes the position of the tracked potential minimum.
No renormalization is applied during propagation.
Electronic relaxation and back-action on the polariton condensate are neglected.

%SSSSSSSSSSSSSSSSS
%SSSSSSSSSSSSSSSSS
%EEEEEEEEEEEEEEEEE
%CCCCCCCCCCCCCCCCC
\subsection*{Simulation parameters}

The simulations use the following excitation parameters:
radial displacement of the pump spots $R=12~\mu$m, 
in-plane pump wavenumber $k_{\mathrm{pump}}=2.5~\mu$m$^{-1}$, 
frequency detunings $\Delta\omega_1=-0.11$~ps$^{-1}$ (unless otherwise stated) and $\Delta\omega_{2,3}=0$,
Gaussian pump width $\sigma=15~\mu$m and 
continuous-wave pump amplitude $F_0=2.4$~meV$\cdot\mu$m$^{-1}$.
The polariton parameters are $m^*=10^{-4}m_e$, 
lifetime $\tau=10$~ps ($\gamma=0.1$~ps$^{-1}$), 
interaction constants $\alpha_1 = -10 \alpha_2=1~\mu$eV$\cdot\mu$m$^2$, and 
the TE--TM splitting constant $\beta=11~\mu$eV$\cdot\mu$m$^2$.
The electronic parameters are $v_F=10^6$~m\,s$^{-1}$ and 
$\Delta_g=30$~meV, corresponding to a band-edge effective mass $m_D=2.6\times10^{-3}m_e$.
The distances between the 2DES and the electron and hole layers of IXs with $\eta_{\text{IX}}=0.8$ are $l_e=1.9$~nm and $l_h=3.5$~nm, respectively, corresponding to a dipole length $d=l_h-l_e=1.6$~nm.
The Coulomb interaction is evaluated using a relative permittivity $\varepsilon\simeq4$, the elementary charge $e=1.602\times10^{-19}$~C, and the vacuum permittivity $\varepsilon_0=8.854\times10^{-12}$~F\,m$^{-1}$.

\section*{Data availability}
All data generated and analyzed during the current study are available from the corresponding author on reasonable request.

\section*{Acknowledgments}
The support of the Ministry of Science and Higher Education of the Russian Federation (No. FSMG-2026-0012),
Saint-Petersburg State University (research grant No.~125022803069-4),
the state assignment in the field of scientific activity of the Ministry of Science and Higher Education of the Russian Federation (theme FZUN-2024-0019, state assignment of VlSU) are acknowledged.

\section*{Conflict of interest}
The authors declare that they have no conflict of interest.

\section*{Authors' contributions}
N.~I.~G. performed simulations, co-developed the theoretical framework, analyzed and interpreted the results, and wrote the manuscript with input from all co-authors.
E.~S.~S. co-developed the theoretical framework, analyzed and interpreted the results, and critically revised the manuscript.
A.~V.~K. initiated the study, developed the concept, and guided the work.
All authors discussed the results and contributed to the manuscript.

%BBBBBBBBBBBBBBBBB
%IIIIIIIIIIIIIIIII
%BBBBBBBBBBBBBBBBB
%LLLLLLLLLLLLLLLLL
\bibliographystyle{naturemag}
\bibliography{main}

\end{document}